\documentclass[conference]{IEEEtran}
\IEEEoverridecommandlockouts
\usepackage{cite}
\usepackage{amsmath,amssymb,amsfonts}
\usepackage{algorithmic}
\usepackage{graphicx}
\usepackage{textcomp}
\usepackage{xcolor}
\usepackage{orcidlink}
\usepackage{balance}
\def\BibTeX{{\rm B\kern-.05em{\sc i\kern-.025em b}\kern-.08em
    T\kern-.1667em\lower.7ex\hbox{E}\kern-.125emX}}

\newcommand{\email}[1]{\texttt{#1}}
\begin{document}

\graphicspath{{./}{figures/}}

\title{Causally Linking Health Application Data and Personal Information Management Tools
\thanks{The authors contributed equally to the paper and their names are listed in alphabetical order.}

}

\author{\IEEEauthorblockN{Saturnino Luz}
\IEEEauthorblockA{\textit{Usher Institute, Edinburgh Medical School} \\
\textit{The University of Edinburgh}\\
Edinburgh, UK \\
\email{s.luz@ed.ac.uk}\\
\orcidlink{https://orcid.org/0000-0001-8430-7875} 0000-0001-8430-7875
}
\and
\IEEEauthorblockN{Masood Masoodian}
\IEEEauthorblockA{\textit{School of Arts, Design and Architecture} \\
\textit{Aalto University}\\
Espoo, Finland \\
\email{masood.masoodian@aalto.fi}\\
\orcidlink{https://orcid.org/0000-0003-3861-6321} 0000-0003-3861-6321
}
}

\maketitle

\begin{abstract}
The proliferation of consumer health devices such as smart watches, sleep monitors, smart scales, etc, in many countries, has not only led to growing interest in health monitoring, but also to the development of a countless number of ``smart'' applications to support the exploration of such data by members of the general public, sometimes with integration into professional health services. While a variety of health data streams has been made available by such devices to users, these streams are often presented as separate time-series visualizations, in which the potential relationships between health variables are not explicitly made visible.  Furthermore, despite the fact that other aspects of life, such as work and social connectivity, have become increasingly digitised, health and well-being applications make little use of the potentially useful contextual information provided by widely used personal information management tools, such as shared calendar and email systems.  This paper presents a framework for the integration of these diverse data sources, analytic and visualization tools, with inference methods and graphical user interfaces to help users by highlighting causal connections among such time-series.
\end{abstract}


\section{Introduction}
\label{sec:introduction}

Healthcare costs are increasing rapidly in most countries due to many
factors, including for instance ageing populations and the increased prevalence of
health conditions that affect older people, as well as other life-style
related health conditions such as obesity -- not to mention
many other physical and mental health and well-being problem related to increased
stress and workload affecting a large percentage of the working population. 
This, combined with a
greater awareness of health and well-being concerns, has led a wider
segment of the general population in many parts of the world to take
more responsibility for managing their own personal health and well-being. To
assist with these personal initiatives, a wide range of consumer
health devices such as smart watches \cite{bib:JatAndGronli2022} and
jewellery \cite{bib:JuAndSpasojevic2015}, sleep monitors, wearable
devices and smart scales, as well as mobile sports applications
\cite{bib:FerrieroEtAl2020} and physiological monitoring tools
\cite{bib:JeongEtAl2022} have become available in recent years, 
often developed by the companies making the actual physical consumer 
devices.

At the same time, it is also becoming more widely acknowledged that
health and well-being are, to a large extent, affected by life factors
such as workload \cite{bib:PaceEtAl2021,bib:HernandezEtAl2021}, social
connectivity and mobility \cite{bib:MariekEtAl2022}.  Despite this
greater awareness, currently there is a disconnection between the use
of health tracking devices, such as those mentioned above, and most
Personal Information Management (PIM) software tools \cite{bib:Lush2014} used
in everyday work and non-work contexts. Yet, these PIM tools,
including for instance email and calendar systems, could provide rich
data sources for measuring workload and other health and well-being
related factors.

In this paper, we present a framework for integrating a diverse range of 
time-based data coming from different sources such as physical devices, 
mobile applications and PIM software, 
to provide input into analytic tools and linked visualizations
that can be used as part of a visual dashboard, called DASH:HEALTH,
for highlighting causal connections among different measured factors
that may affect ones health and well-being.

\section{Linking health and PIM data}
\label{sec:linking-health-pim}

\begin{figure*}[!t]
  \centering
  \includegraphics[width=\linewidth]{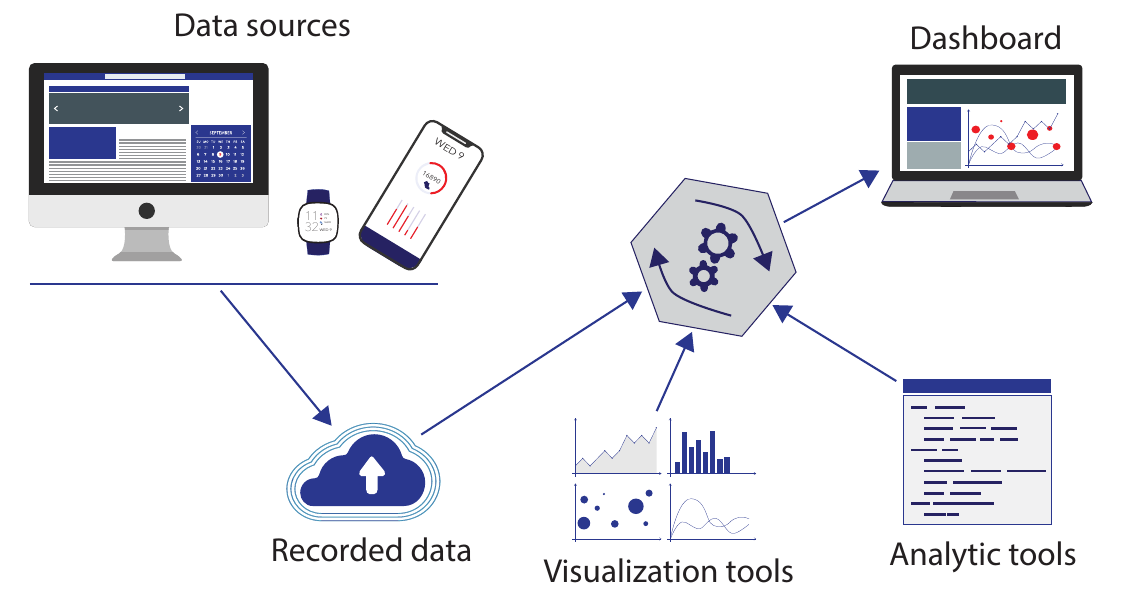}
  \caption{The proposed framework for linking health and PIM data.}
  \label{fig:framework}
\end{figure*}

Figure~\ref{fig:framework} presents a schematic view of our proposed
framework for integrating different types of data, including those
generated by various health-related devices and mobile application, as well
as data that can obtained from PIM software tools. As can be seen, the framework consists of
five main components. These are: 1) data sources, 2) recorded data, 3)
visualization tools, 4) analytic tools, and 5) a dashboard for viewing
and analysing linked health and well-being data, along with the associated PIM data.

\section{Data Sources}
\label{sec:data-sources}

As noted earlier, the aim of our proposed framework is to facilitate
integration of data coming from various sources. Here we are mainly
concerned with data being generated by the aforementioned
health-related devices and mobile applications, as well as data obtained from
PIM software tools. However, the framework could easily be extended in the
future to incorporate a wider range of other kinds of data that are considered to have
an impact on personal health and well-being, including for instance 
food and dietary data, shopping habits, transport data, environmental data,
social networks data, or financial data.
 
In terms of health data, information readily recorded by many consumer
health devices and mobile applications include for instance the following:
\begin{itemize}
\item \textbf{Sleep quality}: with wearable monitoring devices or smart watches.
\item \textbf{Physical activity}: with smart watches or phones. 
\item \textbf{Heart rate}: with smart watches or measurement devices.
\item \textbf{Pulse wave velocity}: using smart scales. 
\item \textbf{Weight}: using smart scales.
\item \textbf{BMI}: using smart scales.
\item \textbf{Fat mass}: using smart scales.
\item \textbf{Muscle mass}: using smart scales.
\item \textbf{Blood pressure}: with electronic consumer devices
  and some smart watches.
\item \textbf{Glucose levels}: with electronic blood sugar monitors.
\end{itemize}

In terms of PIM data, there are not many automated data collection
methods currently being provided by existing PIM software tools. There are,
however, a few proprietary applications and manual workarounds for
extracting data from at least the most commonly used PIM software tools.  
In addition, it is
envisaged that once the importance of such data is realised, in future PIM
software tools are likely to provide methods for exporting, for instance,
workload related data for use by other applications.  At this stage,
it is possible to compile, with some effort, the following PIM data:

\begin{itemize}
\item \textbf{Email volume}: this could included the number of emails
  received, sent, deleted, kept, forwarded, or replied to during
  specific time-periods.
\item \textbf{Calendar appointments}: this could include the number or
  duration of meetings, online or physical, during specific
  time-periods.
\item \textbf{To-do lists}: this could include the number of tasks
  with or without deadlines, and perhaps an indication of the level of
  tasks difficulty or complexity (e.g., see
  \cite{bib:BlomkvistEtAl2004}).
\item \textbf{Communication}: this could include the number of, for
  instance, work-related messages exchanged through social media, or
  phone, audio or video calls.
\end{itemize}

\section{Visualization Tools}
\label{sec:visualization-tools}

Much of the data generated using consumer health devices and mobile
applications is temporal.  While there are differences among such data in
terms of their granularity (i.e., regularity with which the
measurements are made across), they can all be considered as time-series.

Time-series are perhaps the most common form of time-based data, and
as such, there are many different types of visualization methods suitable for
presenting time-series \cite{bib:Tufte2001,bib:AignerEtal2011},
amongst which are various forms of line graphs, bar charts, and area
charts. 
Although, these visualization are easy to understand, for
instance for seeing trends a single time-series dataset, they can be
difficult to use for comparing multiple time-series
\cite{bib:MasoodianEtal2015}. 

Therefore, alternative methods such as \emph{stack
  zooming} \cite{bib:JavedAndElmqvist2010}, and visualizations for combining
heterogeneous temporal data streams \cite{bib:LuzMasoodianAVI04} have been
proposed for making these kinds of comparisons visually possible.
Yet, even these alternatives are perhaps more suited for data expert
than ordinary users, such as those viewing and comparing their own
personal health data.

Data compiled from, or generated by, PIM software tools are not particularly
temporal -- even though all data could be considered temporal in some
sense \cite{bib:AlMohdAli11p}. Therefore, some form of more strict
temporarily must somehow be imposed on such data to make them suitable for
linking with health-related time-series. 

Since the type of linking we
are proposing here is concerned with historical (past) data, rather
than dynamic data, we would argue for adopting a simple time-slicing
rule, which would allow for a kind of bucket sorting of all activities
into different time-periods (e.g., all email received and all meetings
in a given day).  Once this kind of time-slicing has been carried out,
the PIM data could also be visualized using one of the above-mentioned
time-series visualization methods.

\section{Analytic Tools}
\label{sec:analytic-tools}

\begin{figure*}[h!tp]
  \centering
  \includegraphics[width=\linewidth]{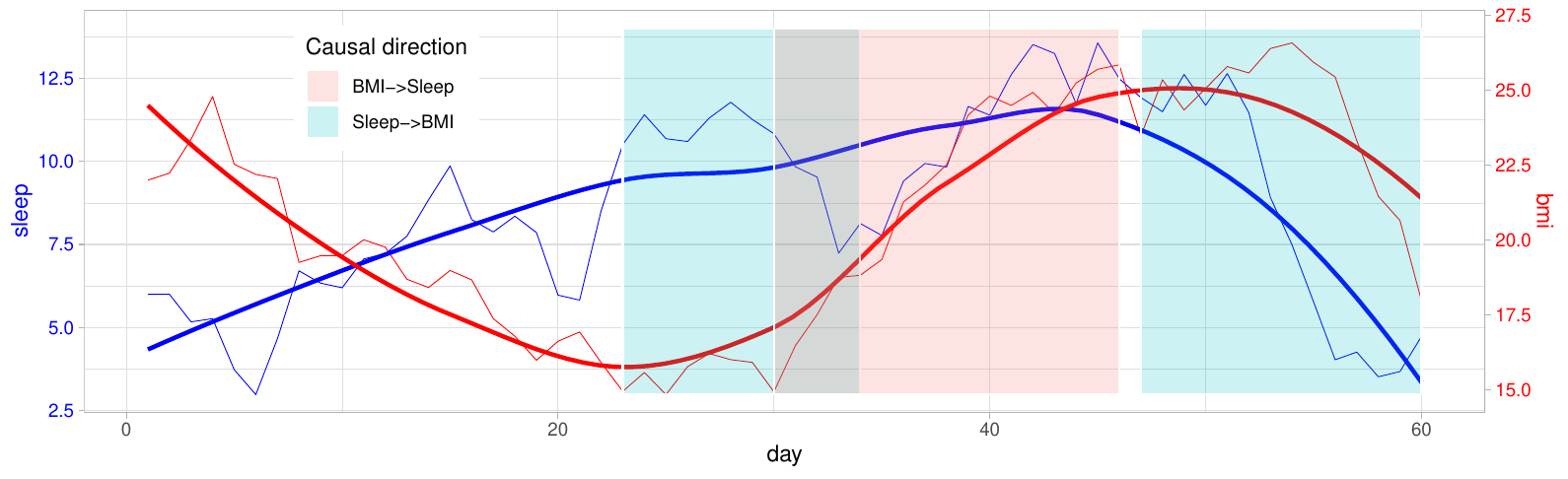}\vspace*{-1ex}
  \caption{Causal discovery on health time-series.}
  \label{fig:cdts}
\end{figure*}

While visualization methods provide various means for users to explore data,
the multi-factorial nature of the combinations of health, activity and
PIM data we envisage might hide connections among these data from
users, overload them with visual complexity, or even mislead them into
seeing connections where they do not exist. To avoid such scenarios,
analytic tools should be provided which can summarise the variables in
question, and point the user to intervals of the time where
correlational or causal links might exist, and suggest further
comparisons to the user to make.

Most health applications display the variables monitored by different
devices (e.g., scales, pedometers, heart rate, etc) over time, often using
standard line graphs along time lines and summary statistics over
selected time periods \cite{bib:ChoeLeeEtAl17un}. PIM software tools, on
the other hand, very rarely rely on time lines, structuring
information along application specific representations of the time
dimension, such as lists ordered by dates for email or planar calendar
layouts for diaries and appointments \cite{bib:AlMohdAli11p}. 

For combinations of health and PIM data, aimed at elucidating possible
connections between events in one's professional and health
biomarkers -- as discussed above -- displaying all information as time
line summaries, can provide natural overviews of activity and
biomarkers, and might help elucidate putative relationships
\cite{bib:FengAgosto19r}.

\subsection{Summary statistics}
\label{sec:summary-statistics}

Summary statistics can be encoded as graphical displays -- the most
common form of user interface presentation for self-tracking data -- or
as text summaries, frequently as annotations to a graph. The most
frequently used visualizations for self-tracking data are line graphs,
bar charts and custom summaries \cite{bib:ChoeLeeEtAl14un}.  While
these representations have been found useful as overviews of
variability in health and activity indicators over time, they are less
useful for exploring relationships among those indicators.

Scatter plots (and often scatter plot matrices) with regression lines
provide effective summaries of correlations, and can be complemented
by results of statistical correlation tests in textual form. One
might, for instance, be able to detect a negative correlation between
Body Mass Index (BMI) and the number of hours slept per night, out of several possible
pairings of health indicators, simply by inspecting a scatter plot
matrix, and suspect that one's lack of sleep might be causing them to
gain weight. However, this presupposes selecting a particular time
interval for correlation analysis, which is an arbitrary
choice. 

Furthermore, the suspicion of a causal connection is only very
weakly supported by evidence of correlation, and the putative causal
direction is a matter of interpretation -- one might postulate,
alternatively, that one's worries about their weight gain might be
causing sleeplessness. In order to address these problems, we propose
complementing these statistical summaries with an analysis of
directional, time-interval sensitive causal effects, as described
below.

\subsection{Causal connections}

Research on causal connections between time-series generated through
activity or biomarker monitoring has gained increased attention in
biomedicine. Monitoring in such settings usually involve taking
measurements from patients on a daily or hourly basis over long
periods of time. This includes data gathered through experimental
designs such as ecological momentary assessments (EMA) and experience
sampling \cite{bib:ShiffmanStoneHufford08ec}, which pose significant
challenges in terms of causal modelling.

The most common approach for modelling the influence of a time-series
on another is Granger Causality \cite{bib:Granger69in}, a technique
that originated in econometrics but has found applications in areas as
diverse as epidemiology and neuroscience
\cite{bib:BresslerSeth11w}. 

While there is criticism of the
interpretation of Granger Causality (GC) as an actual causal relation,
and debate about its meaning in the presence of cross-lagged effects,
GC nevertheless is useful as a first approximation to enable users to
identify and explore relevant connections among the various
time-series generated by health and PIM data. GC therefore forms the
basis of our proposed approach to highlighting areas of interest on
time-series visualizations, where one variable might be influencing
another along the time-line.

Within the GC framework, we model the time-series $X$ (representing,
say, hours of sleep per night) and $Y$ (say, BMI measures) as linear
regression models, where the current value of the variable is
predicted in terms of past values, up to a maximum lag
interval. Intuitively, we say the $X$ (Granger) causes $Y$ if the
including past values of $X$ in the regression model of $Y$ improves
the predictions of current values of $Y$ in relation to a model that
only includes past values of $Y$. Formally, given the model set out in
(\ref{eq:gc}), GC seeks to determine whether the prediction of the
value of variable $y$ at time $t$, (written $y_t$) from $y_{t-l}$
(where $`l`$ is the time lag, or order of the model) is improved by
adding $x_{t-l}$ to the model. If the prediction is improved for all
$t \in T$, where $T$ is the number of time-points considered, we say
that $X$ causes $y$. Prediction improvement can be tested using the
F-test, for instance, with $\gamma = 0$ as the null hypothesis.

\begin{equation}
  \label{eq:gc}
  y_{t} = \alpha \sum_{l=1}^{L} \beta y_{t-l} +  \sum_{l=1}^{L} \gamma x_{t-l} + \epsilon
\end{equation}

While GC analysis alleviates the issue of spurious correlation we
pointed out above in connection with scatter plots, it still does not
address the problem of assessing the connections between time-series
at different time intervals as opposed to the whole series. While this
problem has not been comprehensively addressed in the GC literature,
we adopt the general idea for discovery of GC interval proposed in
\cite{bib:LiZhengEtAl17d}. The basic idea was to test successive time
intervals, varying the lengths of the intervals tested and allowing
for look-ahead to minimise the computational cost of multiple GC
tests. However, to simplify the presentation, here we assume a sliding
window of fixed length $d$ sliding over a pair of time series ($X$ and
$Y$).  For each sub-interval $T'$ thus defined we then model $y_t$
($t \in T'$) as in (\ref{eq:gc}) and test for GC in both directions
(e.g., in one direction BMI corresponds to $Y$ and in the other sleep
corresponds to $Y$). When causality is discovered for a certain
interval we highlight it on the time-line, for example, by overlaying
translucent (alpha-blended) colour coded rectangles onto the regions
where connections were discovered, as shown in Figure~\ref{fig:cdts}.

This figure illustrates a situation where, taken as a whole, BMI does
not cause sleep nor vice-versa. However, in certain intervals (marked
as blue rectangles in Figure~\ref{fig:cdts}), increases in the number
of hours slept cause BMI to decrease, while for a brief interval
(marked in red), the direction of causality is reversed, and as BMI
climbs towards a high value, BMI causes sleep quality to vary, first
shown as a decrease in the number of hours slept, and then as an
increase towards an abnormally high number of hours.

\begin{figure*}[!t]
  \centering
  \includegraphics[width=\linewidth]{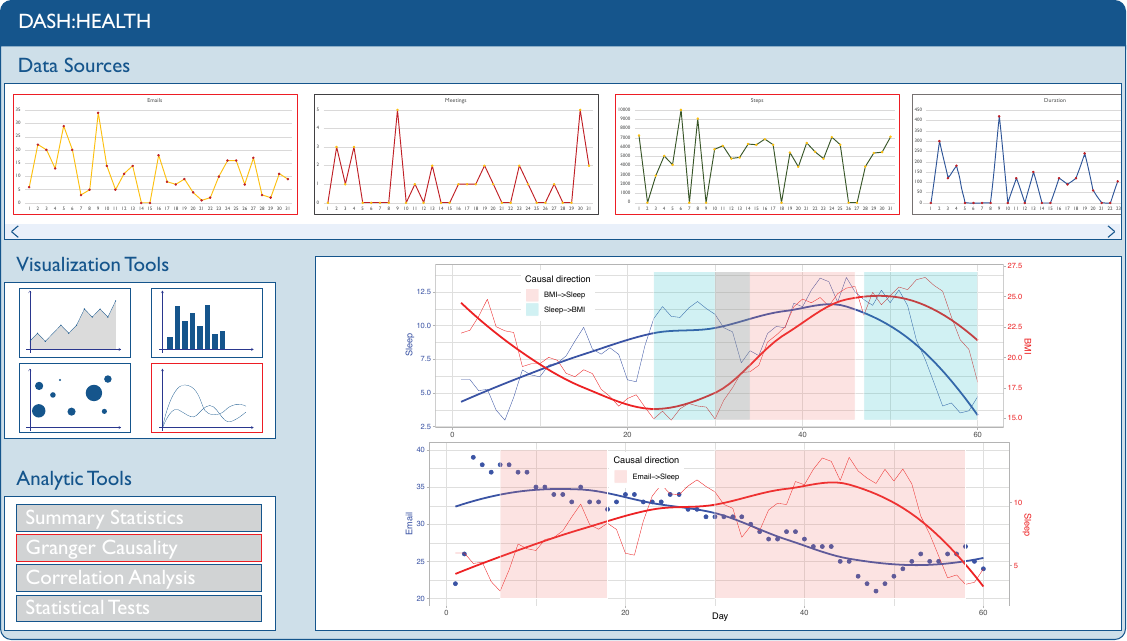}
  \caption{High-fidelity mock-up of DASH:HEALTH user interface.}
  \label{fig:dashboard}
\end{figure*}

Presented with this analysis through the visual summary, the user
might then proceed to investigate possible connections between these
variables and PIM data. The user, might, for example, investigate
whether there is a connection between email volume or meeting
frequency and an increase in BMI, mediated by the effect of the former
on sleep quality, as illustrated in Figure~\ref{fig:dashboard},
bottom-left panel. 

\section{Machine learning enhancements}
\label{sec:mach-learn}

The inference method presented above relies on fairly simply linear
model to infer relationships between two or more time series. However,
there is no guarantee that such relationships are linear. In recent
years new approaches based on kernel methods
\cite{bib:MarinazzoLiaoEtAl11ng} and architectures such as multilayer
perceptrons and recurrent neural networks have been proposed for
inference of Granger causality
\cite{bib:TankCovertEtAl22ngc,bib:YinBarucca22dg}, which can handle
non-linear data.

Outside the Granger causality framework, we have previously employed
machine learning methods to infer relationships among various time
streams in the domain of medical communication
\cite{bib:LuzTOIS12}. These methods assessed interrelationships from a
state change perspective using an underlying transition graph with a
first-order Markovian assumption. For PIM and health marker timeseries
data one could adapt this approach by identifying meaningful ranges of
values both from a clinical (e.g. high blood pressure, low, normal,
and elevated heart rate, low, normal and high BMI, etc) and from a
work activity perspective (high vs normal workload, etc) and thus
discretising them and representing the time series as a sequence of
transitions among such states.

The application of these methods, however, requires larger and more
diverse datasets for model training and validation. We are in the
process of gathering such data.

\section{User Interaction}
\label{sec:user-interaction}
At the heart of the framework proposed here is the graphical user
interface, in the form a visual dashboard, which is meant to support
users with the tasks of visualising, linking, and interacting with
their health and PIM data, to draw accurate conclusions and better
manage their own health and well-being based on their personal data.

As part of a Human-Centred Design process, it is important to
critically evaluate the design of user interfaces and visualizations
before they are fully implemented. We have, therefore, created a
high-fidelity mockup \cite{bib:Sketch2022} of our proposed dashboard
for health, called DASH:HEALTH, which we aim to evaluate through a
formal user study.

Figure~\ref{fig:dashboard} shows the DASH:HEALTH mockup. The top area
of includes all the personal data sources made available to the user
through the framework, viewed using a selected time-series
visualization method (e.g., line graphs). Once two or more data
sources are selected, the user can choose a preferred visualization
tool from a possible set (shown on middle left-hand side) for viewing
the combined time-series (shown on bottom right-hand side). In
addition, the user is able to select an analytics tool (shown on
bottom left-hand side) to identify any important causal effects, as
discussed in the previous section.

\section{Conclusion and Future Work}

In this paper we have proposed and discussed a framework for
monitoring personal health and well-being through the integration of
different types of time-based health and PIM data, and a strategy to
facilitate the visualization and causal analysis of such data. We have
also presented a mockup design for the DASH:HEALTH prototype tool
which could be developed to support users to manage their own personal
health and well-being using such data.

In terms of our future work, we plan to evaluate the DASH:HEALTH
mockup design with health professionals and ordinary users, before
progressing on to developing a functional prototype based on it.  This
type of evaluation-based design and development process aims to result
in greater usability and user adherence to eventual interventions
aimed at improving the user's health and well-being.  This is also
necessary for developing any effective pervasive health and well-being
application.

We are also gathering and curating a data set consisting of
syncronized and combined PIM and personal health data from several
individuals. This, to the best of our knowledge, will be the first
data set of this kind, enabling the research community to explore
relationships between occupational and clinical aspects of health
using AI tools. In addition, we intend to explore more flexible
methods for causal discovery, such as structural equation modelling
and neural network models.

\balance

\bibliographystyle{IEEEtran}
\bibliography{ai4health2023}

\begin{thebibliography}{10}
\providecommand{\url}[1]{#1}
\csname url@samestyle\endcsname
\providecommand{\newblock}{\relax}
\providecommand{\bibinfo}[2]{#2}
\providecommand{\BIBentrySTDinterwordspacing}{\spaceskip=0pt\relax}
\providecommand{\BIBentryALTinterwordstretchfactor}{4}
\providecommand{\BIBentryALTinterwordspacing}{\spaceskip=\fontdimen2\font plus
\BIBentryALTinterwordstretchfactor\fontdimen3\font minus
  \fontdimen4\font\relax}
\providecommand{\BIBforeignlanguage}[2]{{%
\expandafter\ifx\csname l@#1\endcsname\relax
\typeout{** WARNING: IEEEtran.bst: No hyphenation pattern has been}%
\typeout{** loaded for the language `#1'. Using the pattern for}%
\typeout{** the default language instead.}%
\else
\language=\csname l@#1\endcsname
\fi
#2}}
\providecommand{\BIBdecl}{\relax}
\BIBdecl

\bibitem{bib:JatAndGronli2022}
\BIBentryALTinterwordspacing
A.~S. Jat and T.-M. Gr{\o}nli, ``Smart watch for smart health monitoring: A
  literature review,'' in \emph{Bioinformatics and Biomedical Engineering},
  I.~Rojas, O.~Valenzuela, F.~Rojas, L.~J. Herrera, and F.~Ortu{\~{n}}o,
  Eds.\hskip 1em plus 0.5em minus 0.4em\relax Cham: Springer International
  Publishing, 2022, pp. 256--268. [Online]. Available:
  \url{https://doi.org/10.1007/978-3-031-07704-3_21}
\BIBentrySTDinterwordspacing

\bibitem{bib:JuAndSpasojevic2015}
\BIBentryALTinterwordspacing
A.~L. Ju and M.~Spasojevic, ``Smart jewelry: The future of mobile user
  interfaces,'' in \emph{Proceedings of the 2015 Workshop on Future Mobile User
  Interfaces}, ser. FutureMobileUI '15.\hskip 1em plus 0.5em minus 0.4em\relax
  New York, NY, USA: Association for Computing Machinery, 2015, pp. 13--15.
  [Online]. Available: \url{https://doi.org/10.1145/2754633.2754637}
\BIBentrySTDinterwordspacing

\bibitem{bib:FerrieroEtAl2020}
\BIBentryALTinterwordspacing
G.~Ferriero, S.~Vercelli, C.~Fundar\`o, and G.~Ronconi, ``Use of mobile
  applications to collect data in sport, health, and exercise science: A
  narrative review,'' \emph{Journal of Strength and Conditioning Research},
  vol.~34, no.~12, 2020. [Online]. Available:
  \url{https://doi.org/10.1519/JSC.0000000000003365}
\BIBentrySTDinterwordspacing

\bibitem{bib:JeongEtAl2022}
\BIBentryALTinterwordspacing
J.-W. Jeong, W.~Lee, , and Y.-J. Kim, ``A real-time wearable physiological
  monitoring system for home-based healthcare applications,'' \emph{Sensors},
  vol.~22, no.~1, p. 104, 2022. [Online]. Available:
  \url{https://doi.org/10.3390/s22010104}
\BIBentrySTDinterwordspacing

\bibitem{bib:PaceEtAl2021}
\BIBentryALTinterwordspacing
F.~Pace, G.~D'Urso, C.~Zappulla, and U.~Pace, ``The relation between workload
  and personal well-being among university professors,'' \emph{Current
  Psychology}, vol.~40, no.~7, pp. 3417--3424, 2021. [Online]. Available:
  \url{https://doi.org/10.1007/s12144-019-00294-x}
\BIBentrySTDinterwordspacing

\bibitem{bib:HernandezEtAl2021}
\BIBentryALTinterwordspacing
R.~Hernandez, E.~A. Pyatak, C.~L.~P. Vigen, H.~Jin, S.~Schneider,
  D.~Spruijt-Metz, and S.~C. Roll, ``Understanding worker well-being relative
  to high-workload and recovery activities across a whole day: Pilot testing an
  ecological momentary assessment technique.'' \emph{International journal of
  environmental research and public health}, vol.~18, no.~19, 2021. [Online].
  Available: \url{https://doi.org/10.3390/ijerph181910354}
\BIBentrySTDinterwordspacing

\bibitem{bib:MariekEtAl2022}
\BIBentryALTinterwordspacing
M.~M. P.~V. Abeele and M.~H. Nguyen, ``Digital well-being in an age of mobile
  connectivity: An introduction to the special issue,'' \emph{Mobile Media \&
  Communication}, vol.~10, no.~2, pp. 174--189, 2022. [Online]. Available:
  \url{https://doi.org/10.1177/20501579221080899}
\BIBentrySTDinterwordspacing

\bibitem{bib:Lush2014}
\BIBentryALTinterwordspacing
A.~Lush, ``Fundamental personal information management activities –
  organisation, finding and keeping: a literature review,'' \emph{The
  Australian Library Journal}, vol.~63, no.~1, pp. 45--51, 2014. [Online].
  Available: \url{https://doi.org/10.1080/00049670.2013.875452}
\BIBentrySTDinterwordspacing

\bibitem{bib:BlomkvistEtAl2004}
\BIBentryALTinterwordspacing
S.~Blomkvist, I.~Boivie, M.~Masoodian, and J.~Persson, ``From piles to tiles:
  designing for overview and control in case handling systems,'' in
  \emph{Proceedings of the CHISIG Annual Conference on Human-Computer
  Interaction}.\hskip 1em plus 0.5em minus 0.4em\relax Wollongong, Australia:
  Ergonomics Society of Australia, 2004, pp. 161--170. [Online]. Available:
  \url{https://hdl.handle.net/10289/4844}
\BIBentrySTDinterwordspacing

\bibitem{bib:Tufte2001}
E.~R. Tufte, \emph{The Visual Display of Quantitative Information},
  2nd~ed.\hskip 1em plus 0.5em minus 0.4em\relax Graphics Press, 2001.

\bibitem{bib:AignerEtal2011}
W.~Aigner, S.~Miksch, H.~Schumann, and C.~Tominski, \emph{Visualization of
  Time-Oriented Data}.\hskip 1em plus 0.5em minus 0.4em\relax Springer-Verlag,
  2011.

\bibitem{bib:MasoodianEtal2015}
\BIBentryALTinterwordspacing
M.~Masoodian, B.~Endrass, R.~B\"{u}hling, and E.~Andr\'{e}, ``Visualization
  support for comparing energy consumption data,'' in \emph{Proceedings of the
  19th International Conference on Information Visualisation}, ser. IV
  '15.\hskip 1em plus 0.5em minus 0.4em\relax IEEE Computer Society, 2015, pp.
  28--34. [Online]. Available: \url{https://doi.org/10.1109/iV.2015.17}
\BIBentrySTDinterwordspacing

\bibitem{bib:JavedAndElmqvist2010}
\BIBentryALTinterwordspacing
W.~Javed and N.~Elmqvist, ``Stack zooming for multi-focus interaction in
  time-series data visualization,'' in \emph{Proceedings of the IEEE Pacific
  Visualization Symposium}, ser. PacificVis '10.\hskip 1em plus 0.5em minus
  0.4em\relax IEEE Computer Society, 2010, pp. 33--40. [Online]. Available:
  \url{https://doi.org/10.1109/PACIFICVIS.2010.5429613}
\BIBentrySTDinterwordspacing

\bibitem{bib:LuzMasoodianAVI04}
\BIBentryALTinterwordspacing
S.~Luz and M.~Masoodian, ``A mobile system for non-linear access to time-based
  data,'' in \emph{Proceedings of Advanced Visual Interfaces AVI'04}.\hskip 1em
  plus 0.5em minus 0.4em\relax ACM Press, 2004, pp. 454--457. [Online].
  Available: \url{https://dl.acm.org/authorize?717619}
\BIBentrySTDinterwordspacing

\bibitem{bib:AlMohdAli11p}
\BIBentryALTinterwordspacing
M.~R. Al~Nasar, M.~Mohd, and N.~M. Ali, ``Personal information management
  systems and interfaces: An overview,'' in \emph{2011 International Conference
  on Semantic Technology and Information Retrieval}.\hskip 1em plus 0.5em minus
  0.4em\relax IEEE, 2011, pp. 197--202. [Online]. Available:
  \url{https://doi.org/10.1109/STAIR.2011.5995788}
\BIBentrySTDinterwordspacing

\bibitem{bib:ChoeLeeEtAl17un}
\BIBentryALTinterwordspacing
E.~K. Choe, B.~Lee, H.~Zhu, N.~H. Riche, and D.~Baur, ``{Understanding
  self-reflection: how people reflect on personal data through visual data
  exploration},'' in \emph{{PervasiveHealth '17: Proceedings of the 11th EAI
  International Conference on Pervasive Computing Technologies for
  Healthcare}}.\hskip 1em plus 0.5em minus 0.4em\relax New York, NY, USA:
  Association for Computing Machinery, May 2017, pp. 173--182. [Online].
  Available: \url{https://doi.org/10.1145/3154862.3154881}
\BIBentrySTDinterwordspacing

\bibitem{bib:FengAgosto19r}
\BIBentryALTinterwordspacing
Y.~Feng and D.~E. Agosto, ``Revisiting personal information management through
  information practices with activity tracking technology,'' \emph{Journal of
  the Association for Information Science and Technology}, vol.~70, no.~12, pp.
  1352--1367, 2019. [Online]. Available:
  \url{https://doi.org/10.1002/asi.24253}
\BIBentrySTDinterwordspacing

\bibitem{bib:ChoeLeeEtAl14un}
\BIBentryALTinterwordspacing
E.~K. Choe, N.~B. Lee, B.~Lee, W.~Pratt, and J.~A. Kientz, ``Understanding
  quantified-selfers' practices in collecting and exploring personal data,'' in
  \emph{Proceedings of the SIGCHI Conference on Human Factors in Computing
  Systems}, ser. CHI '14.\hskip 1em plus 0.5em minus 0.4em\relax New York, NY,
  USA: Association for Computing Machinery, 2014, pp. 1143--1152. [Online].
  Available: \url{https://doi.org/10.1145/2556288.2557372}
\BIBentrySTDinterwordspacing

\bibitem{bib:ShiffmanStoneHufford08ec}
\BIBentryALTinterwordspacing
S.~Shiffman, A.~A. Stone, and M.~R. Hufford, ``Ecological momentary
  assessment,'' \emph{Annual Review of Clinical Psychology}, vol.~4, pp. 1--32,
  2008. [Online]. Available:
  \url{https://doi.org/10.1146/annurev.clinpsy.3.022806.091415}
\BIBentrySTDinterwordspacing

\bibitem{bib:Granger69in}
\BIBentryALTinterwordspacing
C.~W. Granger, ``Investigating causal relations by econometric models and
  cross-spectral methods,'' \emph{Econometrica}, vol.~37, no.~3, pp. 424--438,
  1969. [Online]. Available: \url{http://www.jstor.org/stable/1912791}
\BIBentrySTDinterwordspacing

\bibitem{bib:BresslerSeth11w}
\BIBentryALTinterwordspacing
S.~L. Bressler and A.~K. Seth, ``Wiener–granger causality: A well established
  methodology,'' \emph{NeuroImage}, vol.~58, no.~2, pp. 323--329, 2011.
  [Online]. Available: \url{https://doi.org/10.1016/j.neuroimage.2010.02.059}
\BIBentrySTDinterwordspacing

\bibitem{bib:LiZhengEtAl17d}
\BIBentryALTinterwordspacing
Z.~Li, G.~Zheng, A.~Agarwal, L.~Xue, and T.~Lauvaux, ``Discovery of causal time
  intervals,'' in \emph{Proceedings of the SIAM International Conference on
  Data Mining (SDM)}, 2017, pp. 804--812. [Online]. Available:
  \url{https://doi.org/10.1137/1.9781611974973.90}
\BIBentrySTDinterwordspacing

\bibitem{bib:MarinazzoLiaoEtAl11ng}
\BIBentryALTinterwordspacing
D.~Marinazzo, W.~Liao, H.~Chen, and S.~Stramaglia, ``Nonlinear connectivity by
  granger causality,'' \emph{NeuroImage}, vol.~58, no.~2, pp. 330--338, 2011.
  [Online]. Available:
  \url{https://www.sciencedirect.com/science/article/pii/S1053811910001382}
\BIBentrySTDinterwordspacing

\bibitem{bib:TankCovertEtAl22ngc}
\BIBentryALTinterwordspacing
A.~Tank, I.~Covert, N.~Foti, A.~Shojaie, and E.~B. Fox, ``Neural granger
  causality,'' \emph{IEEE Transactions on Pattern Analysis and Machine
  Intelligence}, vol.~44, no.~8, pp. 4267--4279, 2022. [Online]. Available:
  \url{https://doi.org/10.1109/TPAMI.2021.3065601}
\BIBentrySTDinterwordspacing

\bibitem{bib:YinBarucca22dg}
\BIBentryALTinterwordspacing
Z.~Yin and P.~Barucca, ``Deep recurrent modelling of granger causality with
  latent confounding,'' \emph{Expert Systems with Applications}, vol. 207, p.
  118036, 2022. [Online]. Available:
  \url{https://doi.org/10.1016/j.eswa.2022.118036}
\BIBentrySTDinterwordspacing

\bibitem{bib:LuzTOIS12}
\BIBentryALTinterwordspacing
S.~Luz, ``The non-verbal structure of patient case discussions in
  multidisciplinary medical team meetings,'' \emph{ACM Transactions on
  Information Systems}, vol.~30, no.~3, pp. 17:1--17:24, 2012. [Online].
  Available: \url{https://dl.acm.org/authorize?6737736}
\BIBentrySTDinterwordspacing

\bibitem{bib:Sketch2022}
\BIBentryALTinterwordspacing
{Sketch B.V.}, ``Wireframe vs mockup vs prototype: What's the difference?''
  April 2022, last visited on 30 March 2023. [Online]. Available:
  \url{https://www.sketch.com/blog/wireframe-vs-mockup-vs-prototype/}
\BIBentrySTDinterwordspacing

\end{thebibliography}

\end{document}